\documentclass[runningheads]{llncs}
\usepackage[T1]{fontenc}
\usepackage{graphicx}

\usepackage{hyperref} % To enable hyperlinks. 
\usepackage{tcolorbox}

\usepackage{pgfplots}
\pgfplotsset{compat=1.18}
\usepackage{pgfplotstable}
\usepackage{subcaption}
\usepackage{rotating}
\usepackage{orcidlink}

\usepackage{booktabs}
\usepackage{caption}
\usepackage{float}
\usepackage{capt-of}
\usepackage{array}
\usepackage{arydshln}
\hypersetup{
    colorlinks=true,
    linkcolor=blue,
    filecolor=magenta,      
    urlcolor=blue,
    pdftitle={Workshop ICSOC24},
    pdfpagemode=FullScreen,
}

\usepackage{ulem} % to use \sout
\begin{document}
\title{From Static to Intelligent: Evolving SaaS Pricing with LLMs}
% \title{Endorsing SaaS Pricings With Intelligence: An LLM-Based Transformation}
\titlerunning{Evolving SaaS Pricing with LLMs}
\author{Francisco Javier Cavero\orcidlink{0009-0004-2453-8814}\and
Juan C. Alonso\orcidlink{0000-0002-1177-9262} \and
Antonio Ruiz-Cortés\orcidlink{0000-0001-9827-1834}}
\authorrunning{F.J. Cavero et al.}
\institute{SCORE Lab, I3US Institute, Universidad de Sevilla, Spain \\
\email{\{fcavero,javalenzuela,aruiz\}@us.es}}
\maketitle              % typeset the header of the contribution
\begin{abstract}

% The Software as a Service (SaaS) model has revolutionized software distribution by offering flexible pricing options to meet diverse customer needs. However, the rapid expansion of the SaaS market has led to a significant increase in pricing complexity, highlighting the need for robust tools to effectively model these structures. This paper tackles the challenges associated with the manual generation and management of SaaS pricing using YAML4SaaS, the Pricing4SaaS metamodel serialization. We propose an LLM-driven approach that automates the generation of YAML4SaaS, enhancing efficiency and consistency while minimizing human error. Our implementation, AI4Pricing2Yaml, features a basic Information Extractor that uses web scraping and Large Language Model (LLM) technologies to accurately extract essential pricing components—plans, features, usage limits, and add-ons—from SaaS websites. Validation against a dataset of 30 distinct commercial SaaS, encompassing over 150 pricings modeled with YAML4SaaS syntax, demonstrates the system's effectiveness in extracting the desired elements across all tasks. However, challenges remain in addressing hallucinations, complex structures, and dynamic content. This work highlights the potential of automating YAML4SaaS generation to streamline SaaS pricing management, offering implications for improved consistency and scalability in an increasingly intricate pricing landscape. Future research will focus on refining extraction capabilities and enhancing the system's adaptability to a wider range of SaaS websites.

The Software as a Service paradigm has revolutionized software distribution by offering flexible pricing options to meet diverse customer needs. 
%
%However, the rapid expansion of the SaaS market has led to increased complexity for DevOps teams, highlighting the need for robust tools to effectively model these pricing structures. This paper accentuates the advantages of having intelligent pricings, facilitating competitive analysis, and enhancing operational decision-making. 
However, the rapid expansion of the SaaS market has introduced significant complexity for DevOps teams, who must manually manage and evolve pricing structures—an approach that is both time-consuming and prone to errors. The absence of automated tools for pricing analysis restricts the ability to efficiently evaluate, optimize, and scale these models. This paper proposes leveraging intelligent pricing (iPricing)—dynamic, machine-readable pricing models—as a solution to these challenges. Intelligent pricing enables competitive analysis, streamlines operational decision-making, and supports continuous pricing evolution in response to market dynamics, leading to improved efficiency and accuracy.
We present an LLM-driven approach that automates the transformation of static HTML pricing into iPricing, significantly improving efficiency and consistency while minimizing human error. Our implementation, AI4Pricing2Yaml, features a basic Information Extractor that uses web scraping and LLMs technologies to extract essential pricing components—plans, features, usage limits, and add-ons—from SaaS websites. Validation against a dataset of 30 distinct commercial SaaS—encompassing over 150 intelligent pricings—demonstrates the system's effectiveness in extracting the desired elements across all steps. However, challenges remain in addressing hallucinations, complex structures, and dynamic content. This work highlights the potential of automating intelligent pricing transformation to streamline SaaS pricing management, offering implications for improved consistency and scalability in an increasingly intricate pricing landscape. Future research will focus on refining extraction capabilities and enhancing the system's adaptability to a wider range of SaaS websites.

\keywords{Software as a Service \and iPricing  \and Large Language Model.}
\end{abstract}
\section{Introduction}

The Software as a Service (SaaS) model has revolutionized software distribution by offering cloud-based access to features and technical support services through recurring subscriptions \cite{saas}. These subscriptions are organized through a \emph{pricing}, i.e., a structure arranged into various plans with optional add-ons. This approach caters to different customer needs and usage levels, allowing users to contract a version of the software based on their requirements and budgets. By offering tailored experiences, SaaS providers can enhance user satisfaction and drive revenue growth through broader market reach.

However, a recent study \cite{ICSOC} reveals a dramatic increase in the \emph{configuration space}, defined as the set of all possible subscription combinations within a pricing. For instance, GitHub's pricing \cite{github} saw an exponential increase (81,354.55\% \cite{ICSOC_LABPACK}) in potential subscription combinations over the last six years, reaching up to 8,960 distinct combinations in 2024 with just 3 plans, 81 features, 9 usage limits, and 14 add-ons. This complexity underscores the urgent need for robust tools to efficiently manage and model SaaS pricing.

Drawing inspiration from \emph{intelligent contracts}—self-adaptive agreements in digital ecosystems—we propose the notion of \emph{intelligent pricing}\footnote{For the sake of brevity, we may also refer to this concept as iPricing.}, i.e. a dynamic, machine-readable pricing that behaves as a software artifact, following the same processes of design, development, and maintenance as other software components. Currently, metamodels like Pricing4SaaS \cite{pricing4saas} lay the groundwork for modeling SaaS pricing, yet manual processes dominate this space. Relying on Natural Intelligence (NI) to model and update SaaS pricing is inefficient and prone to human errors. Furthermore, the lack of standardized guidelines for representing SaaS pricing on websites implies the development of ad hoc solutions for each SaaS, hindering the transformation to iPricing.

This paper introduces an AI-driven approach to automate this transformation. By leveraging LLMs to automate the extraction and modeling of SaaS pricing elements, we reduce manual intervention, minimize errors, and provide a scalable framework. Our solution builds on the Pricing4SaaS metamodel, transforming static pricing information from SaaS websites into an intelligent pricing. As the SaaS configuration space grows exponentially (as seen in \cite{ICSOC}), the adoption of intelligent pricing will become a critical asset for both development and operations teams, facilitating competitive analysis, enhancing operational decision-making, and paving the way for fully automated pricing operations.

In this paper, we present the following key contributions:
\begin{enumerate}
    \item The introduction of the concept intelligent pricing, a promising solution for Pricing-driven Development and Operation of Saas \cite{jcis24}.
    \item A novel AI-driven approach to automate the transformation of iPricing, reducing manual intervention and ensuring consistency across diverse SaaS.
    \item Validation of the previous approach through a working prototype, demonstrating the system's ability to extract and model all the necessary pricing elements across 30 commercial SaaS pricing websites.
    % \item A thorough discussion of the key challenges encountered during the evaluation of this solution, including obstacles in information extraction, managing highly complex pricings, and addressing hallucinations in LLM response. We also propose potential strategies to overcome these barriers in future work.
\end{enumerate}

The remainder of the article is structured as follows. Section \ref{sec:background} provides an overview of SaaS pricings and the role of LLMs in this context. Section \ref{sec:implementation} details our proposed approach and its implementation, while Section \ref{sec:results} analyzes the obtained results and the viability of the approach. Finally, Section \ref{sec:discussion} presents a discussion on the identified challenges and Section \ref{sec:conclusions} presents the conclusions and outlines directions for future work.

\section{Background and Motivation}
\label{sec:background}
\subsection{SaaS Pricing}

In the SaaS deployment model, pricing organizes a service's \emph{features}—distinct attributes that influence subscription decisions—into various plans and optional add-ons. \emph{Plans} determine the baseline level of service, while \emph{add-ons} offer additional functionalities that can be purchased separately, even without a plan \cite{ICSOC}. This flexible approach allows users to tailor their subscriptions to fit specific needs, a key aspect of the SaaS business model. Moreover, the implementation of \emph{usage limits}—which specify the maximum allowable use of certain features before incurring charges or restrictions—help control costs and ensure scalability.

\begin{figure}
    \centering
    \includegraphics[width=\linewidth]{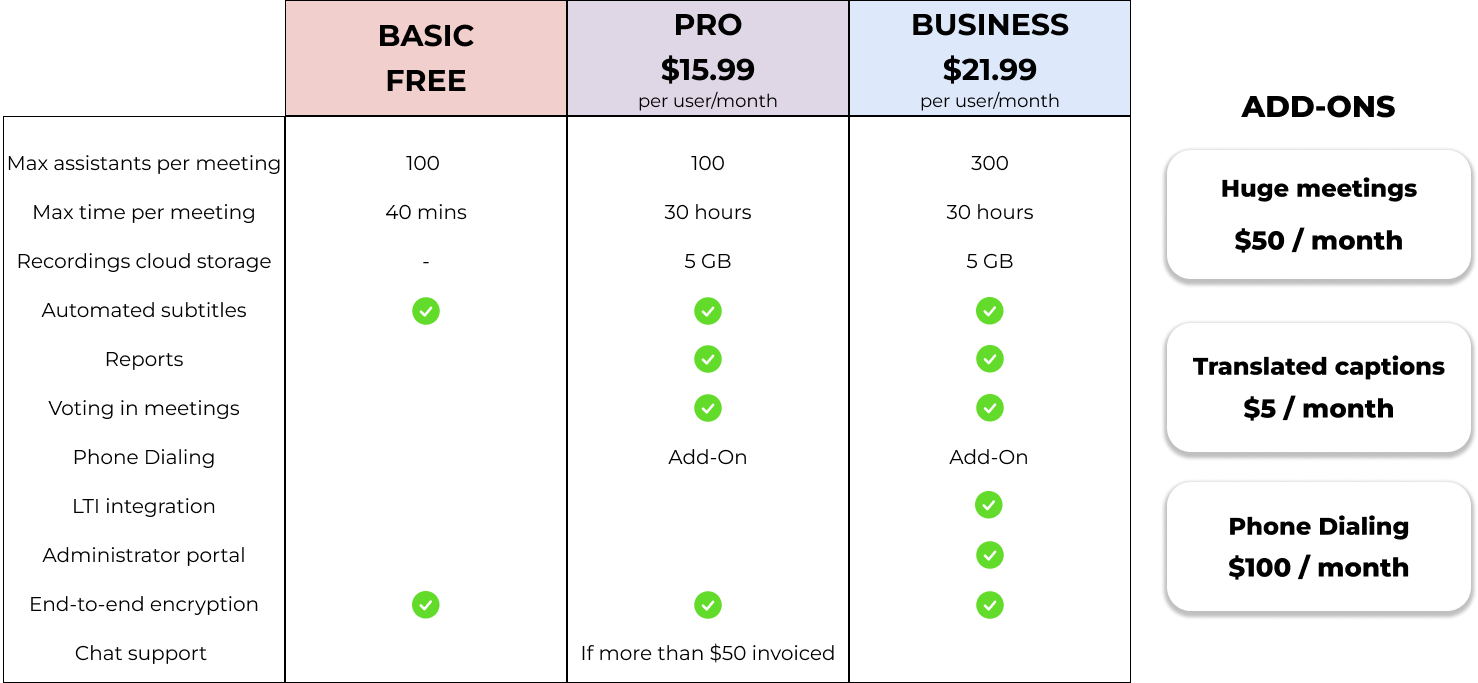}
    \caption{Excerpt of Zoom's pricing. Full pricing is available \href{https://zoom.us/pricing}{here}.}
    \label{fig:zoom}
\end{figure}

Fig. \ref{fig:zoom} illustrates an excerpt of Zoom's pricing \cite{zoom}, showcasing three plans (e.g., Business plan), thirteen features (e.g., access to the administrator portal), three usage limits (e.g., 5 GB of cloud storage for recordings), and four add-ons (e.g., Phone Dialing). Managing the complexity of such pricing smoothly is challenging, as these dynamic structures are subject to frequent revisions to reflect new features, pricing adjustments, or changes in user demand \cite{ICSOC}. These challenges are coined as the \emph{Pricing-driven Development and Operation of SaaS} \cite{jcis24}.

In this context, intelligent pricing serves as the foundation for developing tools that autonomously adapt to changes in SaaS offerings, reducing the need for human intervention, as demonstrated with Pricing4Java and Pricing4React \cite{ICWE24}. This is enabled by a metamodel like Pricing4SaaS \cite{pricing4saas}, which formalizes the relationships between plans, features, usage limits, and add-ons. However, manually converting these complex pricings into machine-readable formats like YAML proves inefficient. As SaaS pricing continues to evolve, the demand for automated tools to transform and manage these iPricings becomes increasingly clear.

% Our work builds on the dataset \cite{ICSOC_LABPACK}, which includes 162 pricing models from 30 commercial SaaS providers over a six-year period (2019–2024). Although manually constructed, this dataset provides a solid foundation for validating AI-driven solutions like ours. Moving forward, automating iPricing modeling will be crucial for streamlining SaaS pricing management, enhancing scalability, and ensuring consistency as the market grows

\subsection{Large Language Models}
% \begin{itemize}
%     \item Brief description of LLMs, their general capabilities, and how they have been applied in natural language processing tasks.
%     \item Examples of how LLMs have been used in other contexts for automating similar tasks, emphasising the time reduction. \aruiz{aquí es donde juan carlos puede darnos algunas referencias de trabajos en los que se genera código en general y esquemas json/yaml en particular}
% \end{itemize}

A \emph{Large Language Model (LLM)} is a sophisticated AI system designed to understand and generate natural language with unprecedented effectiveness \cite{survey}. Models like OpenAI's GPT-4 \cite{gpt4}, Meta's LLaMA 3.1 \cite{llama3}, and Google's Gemini 1.5 \cite{gemini} are trained on vast amounts of text data, enabling them to capture complex patterns, linguistic nuances, and semantic relationships across a wide range of languages and domains. This extensive training equips these LLMs to excel in various tasks, from generating coherent text to multilingual comprehension.

One of the primary factors behind the versatility of LLMs is their architecture, which allows them to process large contexts and retain relevant information over extended text segments \cite{transformers}. This capability is essential for tasks like essay writing or maintaining coherence in long, intricate conversations. In domains such as information extraction (IE), the ability to handle large context windows is crucial for retrieving all the necessary information to generate accurate results.

% A key advancement is the introduction of \emph{LLM agents}, which are advanced AI systems that not only generate text but also autonomously manage complex tasks by interacting with external tools and services. These agents can perform \emph{tool calling}, allowing them to access APIs, databases, or other resources to retrieve real-time information, validate data, or execute specific actions, extending their utility far beyond simple text generation \cite{tool}. Furthermore, LLMs can generate \emph{structured outputs}\footnote{OpenAI gives more info in \href{https://platform.openai.com/docs/guides/structured-outputs/}{here}. Other LLMs also provide this function like \href{https://ai.google.dev/gemini-api/docs/structured-output}{Gemini}.}—such as JSON or YAML—making them particularly effective for domains like IE, where structured outputs facilitate seamless integration into downstream systems.

However, LLMs face significant challenges. One prominent issue is \emph{hallucination}, where models produce text that appears plausible but is factually incorrect or fabricated \cite{hallucinations}. This occurs because LLMs predict the next token based on patterns learned from training data, which doesn't always guarantee factual accuracy. Despite these limitations, LLMs have demonstrated remarkable potential in diverse applications. For instance, in IE domain, they have shown efficacy even in zero-shot scenarios, performing tasks without specific task-related training data \cite{survey}. By optimizing the interaction between the LLM and the prompt (\emph{prompt engineering}), great results can be achieved.

\section{Methodology and Implementation}
\label{sec:implementation}
% \fcavero{Posibilidades para NONAME: AutoYAML4SaaS, AutoPriceYAML, YAMLizeAI, PriceGenAI, Price4SaaSYAMLer u otro. Estoy abierto a sugerencias.}
% \fcavero{Terminar de darle formato y añadir enlaces (como el del repo)}
\subsection{Proposed Automatic Intelligent Pricing Modeler}
\begin{figure}
    \centering
    \includegraphics[width=\textwidth]{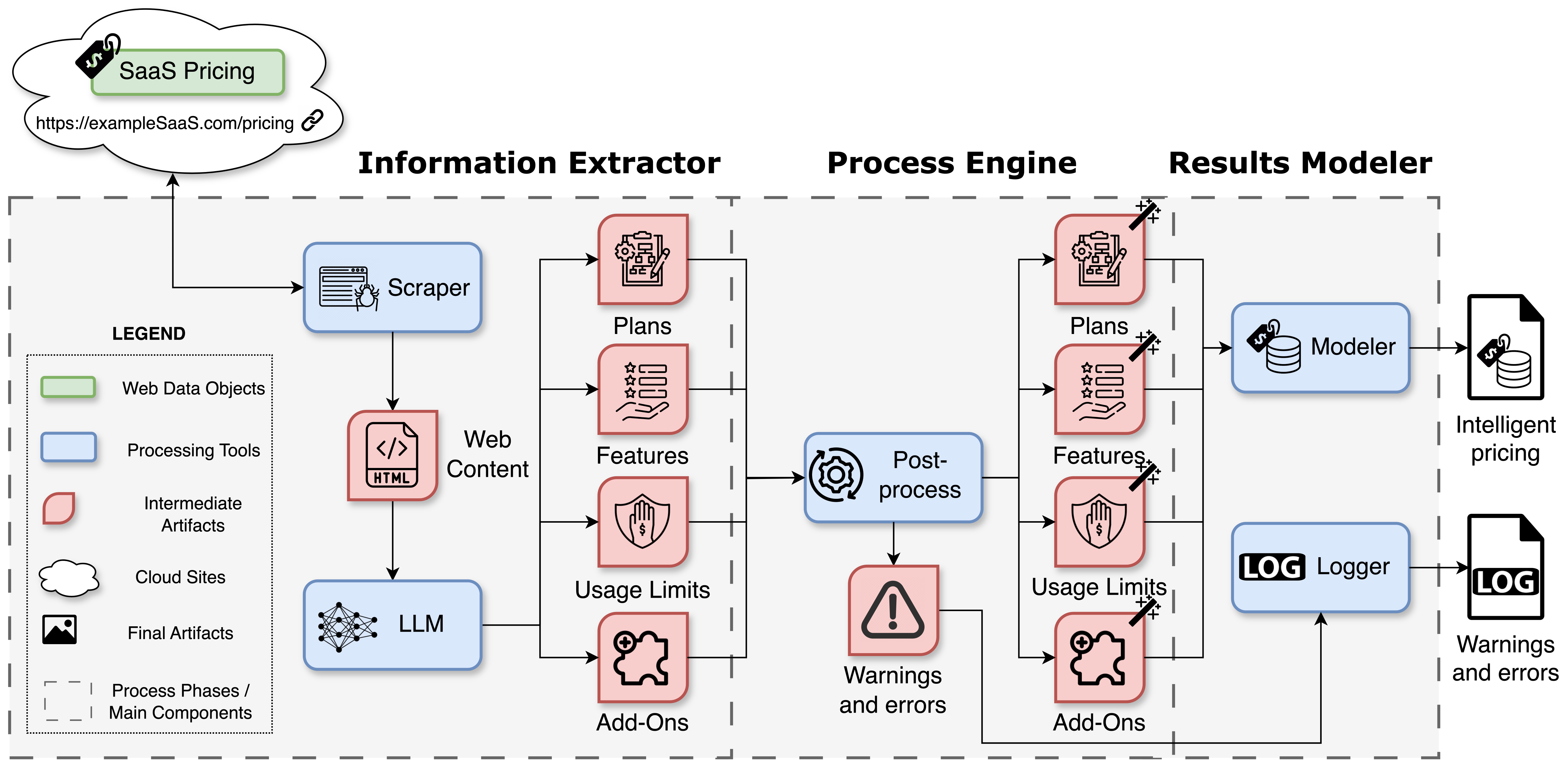}
    \caption{An overview of the proposed automatic intelligent pricing modeler.}
    \label{fig:system}
\end{figure}

% \begin{itemize}
%     \item Presentation of a conceptual schema of how an automatic pricing generator should function, from data collection to YAML4SaaS generation.
%     \item Description of the different (sub)components of the generator, such as data extraction, context interpretation, and conversion to YAML.
% \end{itemize}
This subsection outlines the minimum requirements for developing an extractor capable of automatically transforming a web static pricing into an iPricing by just receiving the URL where the HTML pricing is hosted. As illustrated in Fig. \ref{fig:system}, the final system is divided into three main components:
\begin{itemize}
    \item \emph{Information Extractor:} This component is responsible for retrieving pricing data from a given URL using web scraping technologies. After extracting the data, it leverages an LLM to parse and organize the relevant information, filtering out noise and focusing on essential elements for pricing modeling. These key elements include various plans, features, usage limits, and add-ons.
    \item \emph{Process Engine:} The Process Engine processes the extracted data, validating and verifying the information to correct inconsistencies and mitigate hallucinations commonly found in LLM-generated content (e.g., ensuring there are no duplicate elements). It not only refines the output from the Information Extractor but also generates warnings and errors (e.g., checking discrepancies between annual and monthly pricing) to assist developers in reviewing the resulting YAML file of the subsequent stage.
    \item \emph{Results Modeler:} Finally, the Results Modeler converts the extracted pricing data into a structured file, using a syntax like Pricing2Yaml\footnote{Previously referred % to in the literature 
    as Yaml4SaaS, %this syntax is
    now is rebranded as Pricing2Yaml \cite{pricing2yamlDocs}.} \cite{pricing4saas}, transforming static pricing into an intelligent one. Additionally, it generates a log file with warnings and errors to help identify potential risks and inaccuracies in the LLM-generated content.
\end{itemize}

\subsection{AI4Pricing2Yaml: A Basic Implementation}
\label{sec:implementation:noname}
% \begin{itemize}
%     \item Justification for the choice of tools and LLMs.
%     \item Description of the data flow and anticipated challenges.
% \end{itemize}
% For this implementation, Python was selected due to its popularity and suitability for the project. Python's versatility allows for easy modifications and the potential to offer the system's functionality through an API, making it adaptable to different deployment scenarios, such as a nanoservice.

For this implementation \cite{LABPACK}, Python was selected due to its popularity and suitability for the project. Python's versatility, particularly its ease of integration with LLMs and its rich ecosystem of AI libraries, makes it an ideal choice for rapid development and experimentation. This thriving ecosystem allows for straightforward model invocation, while also providing the potential to offer the system's functionality through an API. This makes Python highly adaptable to different deployment scenarios, such as a nanoservice, which would be the ideal setup for deploying this system.

In our implementation, we employed an existing LLM model. We initially adopted an externally guided approach by applying basic prompt engineering. This approach has yielded surprisingly effective results, as demonstrated in Section \ref{sec:results}. This success suggests that further refinement of the prompts and the use of improved LLM versions could lead to even better outcomes, potentially avoiding the need for fine-tuning. 

% A more detailed discussion on this topic is provided in Section \ref{sec:discussion}.

Considering the three main components of the proposed automatic intelligent pricing modeler, described in the previous subsection, we have fully implemented the first one, the Information Extractor, as it serves as the core component. All the information used by the other components is integrated into the workflow by this component. In other words, the feasibility of the approach largely depends on the performance of the Information Extractor.

The LLM chosen was Gemini 1.5 Flash \cite{gemini}, mainly selected for its context window of up to $10^6$ tokens, allowing the entire HTML content of the SaaS pricing page to be fully processed. It also features a free tier, making it ideal for research without incurring costs. Additionally, this LLM is available through an API, with a pay-as-you-go system for the paid tier (Model as a Service).

For extracting pricing information, Selenium was utilized. Selenium's ability to handle JavaScript-rendered content provides an advantage over more traditional tools like BeautifulSoup. This capability allows web scripts to be processed by the LLM. However, dynamic content, such as elements that require user interaction (e.g., clicking a button or opening a modal), can pose challenges in extracting all relevant information accurately.

Finally, an important consideration for the AI4Pricing2Yaml implementation is its design to extract information from a single webpage. If the webpage does not contain all the necessary information, it may be challenging (not to say impossible) to model accurate pricing. Additionally, the current version of the algorithm is optimized for URLs that include a comparison table with all plans and features info. Applying it to different-structured SaaS webpages may result in unexpected outcomes in both execution and the result files.

\section{Results and Validation}
\label{sec:results}
% \begin{itemize}
%     \item Description of the García et al. dataset, including why it was chosen as the ground truth for validating NONAME.
%     \item Explanation of the methods used to validate NONAME’s accuracy and coverage, including metrics and comparative tests.
%     \item Quantification of the percentage of pricing elements that NONAME was able to generate automatically.
%     \item Evaluation of NONAME’s accuracy in generating YAML4SaaS compared to the García et al. ground truth.
%     \item Validation of the García et al. dataset and confirmation that it remains accurate.
%     \item Identification of NONAME’s limitations, such as potential errors in data interpretation or inconsistencies in YAML generation.
% \end{itemize}

In this section, we evaluate the performance of the implemented Information Extractor using the dataset from \cite{ICSOC_LABPACK}, which contains pricings from 30 commercial SaaS to assess reliability across multiple domains. The evaluation focuses on extracting plans, features, usage limits, and add-ons. The supplementary material \cite{LABPACK} contains an in-depth breakdown of the results.

\subsection{Methodology}
We utilized the implementation detailed in Section \ref{sec:implementation:noname} across the dataset's SaaS to extract various metrics and evaluate the system’s performance \cite{LABPACK}. For each SaaS, we extracted their pricing and categorized the results as follows:

\begin{itemize}
    \item \emph{True Positives (TP):} Accurate extractions of relevant information, such as pricing features, that were successfully retrieved.
    \item \emph{False Positives (FP):} Instances of hallucination where the system inaccurately returned incorrect or non-existent data, such as confusing add-ons with plans or fabricating features.
    \item \emph{False Negatives (FN):} Relevant data that should have been extracted but was overlooked, despite being already scraped (e.g., missing features).
    \item \emph{True Negatives (TN):} Data not extracted because it was absent from the scraped information, although it might be accessible through dynamic interactions (e.g., by clicking a button to load and display additional data).
\end{itemize}

We developed a point-based scoring system to measure extraction accuracy. A full point is awarded for each correct extraction (TP or TN) and for incorrect extractions (FP or FN). Partial or incomplete extractions are handled by assigning half points; for example, if a feature is correctly extracted but misrepresented across plans (e.g., falsely marked as available in more plans than it actually is), half a point is awarded to both the TP and FP categories. This system provides a nuanced evaluation of performance, balancing correct and incorrect extractions across diverse SaaS pricings. The exact results are provided in Table \ref{tab:results}.

Additionally, three key metrics were applied to all the tables where the extraction tasks were performed: \emph{accuracy}—the percentage of correct extractions (TP and TN) relative to the total number of extractions—, \emph{precision}—the percentage of correct positive extractions (TP) out of all positive extractions (TP and FP)—, and \emph{recall}—the percentage of correct positive extractions (TP) out of the total expected positive extractions (TP and FN). Among these metrics, recall is the most important for us, as it validates the viability of the approach by demonstrating that extracting all relevant information is achievable.

Notably, only 15 out of the 30 SaaS platforms had their features, usage limits, and add-ons successfully extracted. This limitation is primarily due to the absence of structured HTML tables or reliance on dynamically generated content, which hindered the extraction process.

\setlength{\tabcolsep}{1pt} % Ajusta la separación entre columnas
\begin{sidewaystable}
\centering
\caption{Extraction Metrics for SaaS (Plans, Features, Usage Limits, Add-Ons)}
\vspace{-4mm}
\resizebox{1.0\textwidth}{!}{
\begin{tabular}{lcccccccc|cccccccc|cccccccc|cccccccc}
\toprule
\multicolumn{1}{c}{} & \multicolumn{8}{c}{\textbf{Plans}} & \multicolumn{8}{c}{\textbf{Features}} & \multicolumn{8}{c}{\textbf{Usage Limits}} & \multicolumn{8}{c}{\textbf{Add-Ons}} \\
\cmidrule(l){2-9} \cmidrule(l){10-17} \cmidrule(l){18-25} \cmidrule(l){26-33}
\textbf{SaaS} & \textbf{TP} & \textbf{FP} & \textbf{FN} & \textbf{TN} & \textbf{T} & \textbf{A (\%)} & \textbf{P (\%)} & \textbf{R (\%)} & \textbf{TP} & \textbf{FP} & \textbf{FN} & \textbf{TN} & \textbf{T} & \textbf{A (\%)} & \textbf{P (\%)} & \textbf{R (\%)} & \textbf{TP} & \textbf{FP} & \textbf{FN} & \textbf{TN} & \textbf{T} & \textbf{A (\%)} & \textbf{P (\%)} & \textbf{R (\%)} & \textbf{TP} & \textbf{FP} & \textbf{FN} & \textbf{TN} & \textbf{T} & \textbf{A (\%)} & \textbf{P (\%)} & \textbf{R (\%)} \\ 
\midrule
\textbf{Box} & 7 & 0 & 0 & 0 & 7 & 100 & 100 & 100 & & & & & & & & & & & & & & & & & & & & & & & & \\ \hdashline
\textbf{Buffer} & 2 & 0 & 0 & 2 & 4 & 100 & 100 & 100 & 64 & 1 & 0 & 0 & 65 & 98.5 & 98.5 & 100 & 4.5 & 1 & 2.5 & 0 & 8 & 56.3 & 81.8 & 64.3 & 1 & 1 & 0 & 0 & 2 & 50 & 50 & 100 \\ \hdashline
\textbf{Canva} & 2 & 0 & 0 & 2 & 4 & 100 & 100 & 100 & & & & & & & & & & & & & & & & & & & & & & & &  \\ \hdashline
\textbf{Clickup} & 2 & 0 & 0 & 2 & 4 & 100 & 100 & 100 & & & & & & & & & & & & & & & & & & & & & & & & \\ \hdashline
\textbf{Clockify} & 8 & 3 & 0 & 0 & 11 & 72.7 & 72.7 & 100 & 68 & 0 & 0 & 0 & 68 & 100 & 100 & 100 & 1.5 & 0.5 & 1 & 0 & 3 & 50 & 75 & 60 & 0 & 1 & 0 & 0 & 1 & 0 & 0 & - \\ \hdashline
\textbf{Crowdcast} & 2 & 1 & 0 & 0 & 3 & 66.7 & 66.7 & 100 & & & & & & & & & & & & & & & & & & & & & & & & \\ \hdashline
\textbf{Databox} & 3 & 2 & 0 & 0 & 5 & 60 & 60 & 100 & & & & & & & & & & & & & & & & & & & & & & & & \\ \hdashline
\textbf{Deskera} & 3 & 0 & 0 & 0 & 3 & 100 & 100 & 100 & & & & & & & & & & & & & & & & & & & & & & & & \\ \hdashline
\textbf{Dropbox} & 3.5 & 0.5 & 0 & 2 & 6 & 91.7 & 87.5 & 100 & 61 & 16 & 1 & 0 & 78 & 78.2 & 79.2 & 98.4 & 8 & 0.5 & 4.5 & 0 & 13 & 61.5 & 94.1 & 64 & 1 & 0 & 0 & 0 & 1 & 100 & 100 & 100 \\ \hdashline
\textbf{Evernote} & 2 & 2 & 0 & 0 & 4 & 50 & 50 & 100 & 28.5 & 0 & 50 & 0 & 29 & 98.3 & 100 & 98.3 & 6 & 0 & 0 & 0 & 6 & 100 & 100 & 100 & 0.5 & 3 & 0.5 & 0 & 4 & 12.5 & 14.3 & 50 \\ \hdashline
\textbf{Figma} & 3.5 & 4.5 & 0 & 0 & 8 & 43.8 & 43.8 & 100 & 88.5 & 0.5 & 0 & 0 & 89 & 99.4 & 99.4 & 100 & 1.5 & 0 & 0.5 & 0 & 2 & 75 & 100 & 75 & 0 & 0 & 0 & 0 & 0 & 100 & 100 & 100 \\ \hdashline
\textbf{GitHub} & 1.5 & 5.5 & 0 & 0 & 7 & 21.4 & 21.4 & 100 & & & & & & & & & & & & & & & & & & & & & & & & \\ \hdashline
\textbf{Hypercontext} & 2 & 2 & 0 & 0 & 4 & 50 & 50 & 100 & 52 & 0 & 0 & 0 & 52 & 100 & 100 & 100 & 0 & 0 & 0 & 0 & 0 & 100 & 100 & 100 & 0 & 0 & 0 & 0 & 0 & 100 & 100 & 100 \\ \hdashline
\textbf{Jira} & 2 & 1 & 0 & 1 & 4 & 75 & 66.7 & 100 & 38 & 10.5 & 9.5 & 0 & 58 & 65.5 & 78.4 & 80 & 5 & 0 & 0 & 0 & 5 & 100 & 100 & 100 & 0 & 20 & 1 & 0 & 21 & 0 & 0 & 0 \\ \hdashline
\textbf{Mailchimp} & 4 & 0 & 0 & 0 & 4 & 100 & 100 & 100 & 76.5 & 26 & 3.5 & 0 & 106 & 72.2 & 74.6 & 95.6 & 6 & 0 & 1 & 0 & 7 & 85.7 & 100 & 85.7 & 2 & 1 & 0 & 0 & 3 & 66.7 & 66.7 & 100 \\ \hdashline
\textbf{Microsoft 365} & 2 & 0 & 0 & 2 & 4 & 100 & 100 & 100 & 39.5 & 0 & 5.5 & 0 & 45 & 87.8 & 100 & 87.8 & 3.5 & 23.5 & 1 & 0 & 28 & 12.5 & 13 & 77.8 & 1 & 4 & 0 & 0 & 5 & 20 & 20 & 100 \\ \hdashline
\textbf{Notion} & 2 & 3 & 0 & 1 & 6 & 50 & 40 & 100 & & & & & & & & & & & & & & & & & & & & & & & & \\ \hdashline
\textbf{Openphone} & 3 & 3 & 0 & 0 & 6 & 50 & 50 & 100 & & & & & & & & & & & & & & & & & & & & & & & & \\ \hdashline
\textbf{Overleaf} & 7 & 0 & 0 & 0 & 7 & 100 & 100 & 100 & 7.5 & 4.5 & 0 & 0 & 12 & 62.5 & 62.5 & 100 & 2 & 0 & 1 & 0 & 3 & 66.7 & 100 & 66.7 & 0 & 3 & 0 & 0 & 3 & 0 & 0 & - \\ \hdashline
\textbf{Planable} & 3 & 1 & 0 & 0 & 4 & 75 & 75 & 100 & & & & & & & & & & & & & & & & & & & & & & & & \\ \hdashline
\textbf{Postman} & 4 & 9 & 0 & 0 & 13 & 30.8 & 30.8 & 100 & 72 & 3.5 & 2.5 & 0 & 78 & 92.3 & 95.4 & 96.6 & 8 & 0 & 1 & 0 & 9 & 88.9 & 100 & 88.9 & 11 & 1 & 3 & 0 & 15 & 73.3 & 91.7 & 78.6 \\ \hdashline
\textbf{Pumble} & 5 & 0 & 0 & 0 & 5 & 100 & 100 & 100 & & & & & & & & & & & & & & & & & & & & & & & & \\ \hdashline
\textbf{Quip} & 3 & 0 & 0 & 0 & 3 & 100 & 100 & 100 & 10 & 0 & 0 & 0 & 10 & 100 & 100 & 100 & 0 & 0 & 0 & 0 & 0 & 100 & 100 & 100 & 0 & 0 & 0 & 0 & 0 & 100 & 100 & 100 \\ \hdashline
\textbf{Salesforce} & 2.5 & 13.5 & 0 & 0 & 16 & 15.6 & 15.6 & 100 & & & & & & & & & & & & & & & & & & & & & & & & \\ \hdashline
\textbf{Slack} & 4 & 0 & 0 & 0 & 4 & 100 & 100 & 100 & 38.5 & 2 & 3.5 & 0 & 44 & 87.5 & 95.1 & 91.7 & 7.5 & 0.5 & 2 & 0 & 10 & 75 & 93.8 & 78.9 & 4 & 1 & 0 & 0 & 5 & 80 & 80 & 100 \\ \hdashline
\textbf{Tableau} & 3 & 4 & 0 & 0 & 7 & 42.9 & 42.9 & 100 & 40 & 0 & 0 & 0 & 40 & 100 & 100 & 100 & 0 & 0 & 0 & 0 & 0 & 100 & 100 & 100 & 3 & 0 & 3 & 0 & 6 & 50 & 100 & 50 \\ \hdashline
\textbf{Trustmary} & 2.5 & 7 & 0 & 1.5 & 11 & 36.4 & 26.3 & 100 & & & & & & & & & & & & & & & & & & & & & & & & \\ \hdashline
\textbf{UserGuiding} & 3 & 0 & 0 & 0 & 3 & 100 & 100 & 100 & & & & & & & & & & & & & & & & & & & & & & & & \\ \hdashline
\textbf{Wrike} & 5 & 0 & 0 & 0 & 5 & 100 & 100 & 100 & 78 & 3 & 0 & 0 & 81 & 96.3 & 96.3 & 100 & 2 & 0 & 2 & 0 & 5 & 40 & 100 & 40 & 3 & 2 & 1 & 0 & 6 & 50 & 60 & 75 \\ \hdashline
\textbf{Zapier} & 1.5 & 9.5 & 0 & 2 & 13 & 26.9 & 13.6 & 100 & & & & & & & & & & & & & & & & & & & & & & & & \\ \midrule
\textbf{Mean} & & & & & & \textbf{64.3} & \textbf{61.4} & \textbf{100} & & & & & & \textbf{88.2} & \textbf{91.1} & \textbf{96.4} & & & & & & \textbf{67} & \textbf{83.8} & \textbf{77.8} & & & & & & \textbf{53.5} & \textbf{63} & \textbf{81} \\
\textbf{Median} & & & & & & \textbf{75} & \textbf{73.9} & \textbf{100} & & & & & & \textbf{96.3} & \textbf{98.5} & \textbf{100} & & & & & & \textbf{75} & \textbf{100} & \textbf{78.9} & & & & & & \textbf{50} & \textbf{73.3} & \textbf{100} \\
\bottomrule
\end{tabular}
}
\label{tab:results}
\end{sidewaystable}

% This issue, along with potential solutions, will be discussed in detail in Section \ref{sec:discussion}.

\subsection{Plans Extraction}

Our LLM-based system achieved 100\% accuracy in extracting SaaS pricing plans for 13 out of 30 SaaS providers. However, the overall mean accuracy was 64.3\%, with a median of 75\%. Key challenges included mistakenly identifying add-ons as plans and extracting plans from other products when multiple offerings appeared on the same pricing page, as observed with Figma and FigJam \cite{figma}.

Despite this, the system achieved perfect recall, successfully extracting all plans. SaaS like Box, Buffer, Canva, and ClickUp achieved 100\% precision and recall, while others like Salesforce (15.6\%) and Zapier (26.9\%) had lower accuracy due to more FP, making it harder to distinguish plans from other elements.

Additionally, the system struggled with extracting dynamically loaded prices, which only become visible after user interactions, such as clicking a button.

\subsection{Features Extraction}

Our feature extraction system performed well, with a mean accuracy of 88.2\% and a median of 96.3\% across tested SaaS. Precision reached 91.1\%, and recall was 96.4\%, showing effective extraction with minimal hallucinations.

Several SaaS achieved perfect precision and recall, while Buffer and Figma also had near-perfect accuracy (98.5\% and 99.4\%). However, challenges arose with SaaS like Dropbox (16 FP out of 78 features) and Mailchimp (26 FP out of 106), where misclassifications and hallucinations led to false positives. 

In contrast, Jira's recall was 80\% due to missed features caused by the model skipping rows in HTML tables—an issue requiring further investigation.

\subsection{Usage Limits Extraction}

In the usage limits extraction, system performance varied, with a mean accuracy of 67\% and a median of 75\%. Despite challenges, the model achieved a mean precision of 83.8\% and recall of 77.8\%.

SaaS like Evernote, Jira, and Postman performed flawlessly, with 100\% recall and accuracy, while Mailchimp and Slack also showed strong results (85.7\% and 78.9\% recall, respectively). However, SaaS like Buffer (56.3\% accuracy) and Clockify (50\%) struggled with missed limits and hallucinations, while Microsoft 365 had the worst performance, at 12.5\% accuracy and 13\% precision.

The main challenge is interpreting and condensing usage limit information, further complicated by the varied presentation styles across SaaS websites.

\subsection{Add-Ons Extraction}

The add-ons extraction showed mixed results, with a mean accuracy of 53.5\% and a median of 50\%. Despite lower accuracy, the system achieved a strong mean recall of 81\%, though precision was 63\%.

High performers included Dropbox and Slack, both achieving perfect recall, while Postman reached 73.3\% accuracy and 91.7\% precision. Mailchimp also performed well, with 66.7\% precision and recall.

However, the system struggled with Jira, Clockify, Overleaf, and Evernote. Jira, Clockify, and Overleaf had no true positives, and Overleaf and Clockify—two SaaS without add-ons—incorrectly extracted non-existent add-ons, rendering recall unmeasurable.

Two prompts were employed: one to extract add-ons alongside usage limits within feature tables, as seen in \href{https://buffer.com/pricing}{Buffer's pricing}, and another to extract the remaining add-ons directly from the HTML, which is the more common approach, as used by \href{https://www.tableau.com/pricing}{Tableau's pricing}.

\subsection{Ideal Extraction}
SaaS pricing is often presented in unstructured formats, making automated extraction difficult. Based on our vision of results and a six-year analysis of SaaS pricing webpages \cite{ICSOC}, we propose an ``ideal SaaS pricing webpage'' with three clearly defined sections:

\begin{enumerate}
    \item \emph{Plans:} Clearly structured subscription options, as in \href{https://userguiding.com/pricing}{UserGuiding's pricing}.
    \item \emph{Comparison table:} Features should be presented in a structured HTML table, preferably as boolean values without any unnecessary details or complexities (e.g., \href{https://quip.com/about/pricing}{Quip's pricing}).
    \item \emph{Add-ons:} A structured add-ons section, detailing each add-on's name, price, and unit, and specifying required plans (e.g., \href{https://www.postman.com/pricing/}{Postman's pricing}).
\end{enumerate}

Usage limits should follow a consistent format, like \href{https://evernote.com/compare-plans}{Evernote's pricing}, and appear within the features table. Add-on-specific features should be listed in a separate table, without overlapping with the main features table. All information should be static, requiring no extra actions to view complete details. This structure will enhance clarity, facilitate extraction, and streamline future transformations, though it may not ensure perfect results.

\begin{figure}[h!]
\centering
\begin{tikzpicture}
  \centering
  \begin{axis}[
        ybar, axis on top,
        width=\textwidth,
        height=0.5\textwidth,
        bar width=0.4cm,
        ymin=0, ymax=100,
        axis x line*=bottom,
        axis y line*=right,
        y axis line style={opacity=0},
        tickwidth=0pt,
        enlarge x limits=true,
        % legend style={
        %     at={(0.025,1.1)},
        %     anchor=north,
        %     legend columns=1,
        %     /tikz/every even column/.append style={column sep=0.5cm}
        % },
        legend style={
            at={(0.5,-0.15)},
            anchor=north,
            legend columns=-1,
            /tikz/every even column/.append style={column sep=0.5cm}
        },
        ylabel={Percentage (\%)},
        symbolic x coords={
           Plans, Features, Usage Limits, Add-Ons},
       xtick=data,
       nodes near coords={
        \pgfmathprintnumber[precision=1]{\pgfplotspointmeta}
       }
    ]
    \addplot [draw=none, fill=blue!30, bar shift=-0.7cm] coordinates {
      (Plans,64.3)
      (Features, 88.2) 
      (Usage Limits,67)
      (Add-Ons,53.5) };
   \addplot [draw=none,fill=red!30, bar shift=0cm] coordinates {
      (Plans,61.4)
      (Features,91.1) 
      (Usage Limits,83.8)
      (Add-Ons,63) };
   \addplot [draw=none, fill=green!30, bar shift=0.7cm] coordinates {
      (Plans,100)
      (Features,96.4) 
      (Usage Limits,77.8)
      (Add-Ons,81) };

    \legend{Accuracy, Precision, Recall}
  \end{axis}
\end{tikzpicture}

\captionsetup{width=\linewidth}
\caption{Summary of accuracy, precision, and recall metrics per extracted element.}
\label{fig:summary}
\end{figure}
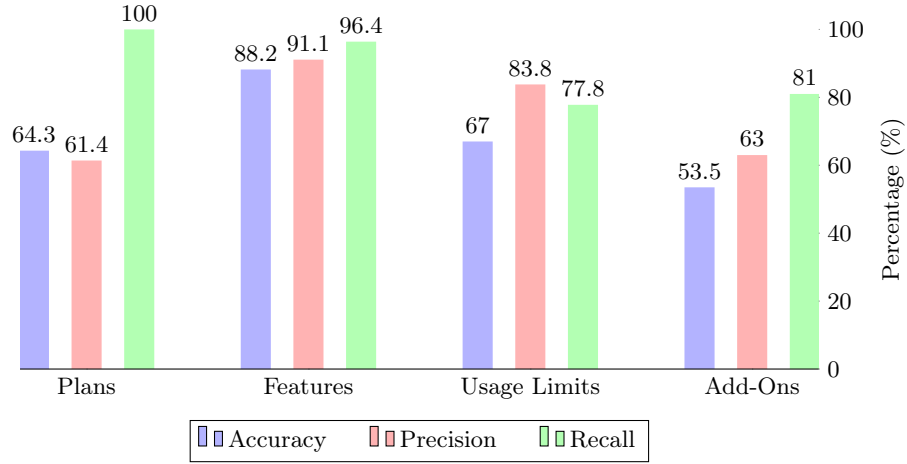

\vspace{-10pt}
\section{Discussion}
\label{sec:discussion}
As shown in Section \ref{sec:results} the extraction system performs well across multiple SaaS, effectively capturing key pricing elements. High mean recall values, along with strong accuracy and precision (see Fig. \ref{fig:summary}), highlight the system’s potential for effectively modeling plans, features, usage limits, and add-ons from a static HTML pricing. However, several challenges persist, especially when dealing with dynamic content, complex usage limit structures, add-ons, and pages featuring multiple products. These challenges can be grouped into two categories:

\begin{itemize}
    \item \emph{Internal:} These challenges stem from the system’s difficulties in accurately modeling extracted elements. Issues such as hallucinations, incorrect modeling of usage limits and add-ons, and occasionally missing features fall under this category. To address these issues, we propose exploring more advanced models (e.g., Gemini 1.5 Pro \cite{gemini}), enhancing prompt engineering, and leveraging advanced functions like tool calling \cite{geminiFunctions} or structured outputs \cite{geminiOutput}.
    \item \emph{External:} These challenges relate to the data extraction process itself, such as difficulties in recognizing tables without HTML tags or interacting with dynamic content (e.g., requiring clicks to load additional data). Implementing an LLM agent could significantly improve this, allowing the system to intelligently navigate and interact with complex or hidden content, ensuring comprehensive and accurate data extraction while minimizing noise. Moreover, the integration of knowledge graphs could provide a more structured approach to capturing relationships between pricing elements, enabling to reduce the noise and size of the original HTML content.
\end{itemize}

In summary, addressing both internal and external challenges will lead to substantial improvements in the system’s output. While internal solutions focus on enhancing the system's processing and modeling capabilities, external solutions aim to optimize the quality of input data and reduce its complexity. %Therefore, enhancing the underlying LLM capabilities, combined with better data handling and interaction strategies, will enable a more efficient and reliable transformation of SaaS pricing into an intelligent one.

\section{Conclusions and Future Work}
\label{sec:conclusions}
% \begin{itemize}
%     \item Synthesis of the study’s main findings, focusing on the feasibility, accuracy, and applicability of NONAME.
%     \item Reflection on the potential impact of automating YAML4SaaS generation on the SaaS ecosystem.
%     \item Proposals for future research and development, including improving NONAME and expanding its capabilities.
% \end{itemize}

This study successfully demonstrates AI4Pricing2Yaml feasibility in automating intelligent pricing transformation, addressing its complexities. Our findings show that the system efficiently extracts key components—plans, features, usage limits, and add-ons—from various commercial SaaS offerings, achieving high accuracy and recall. Notably, feature extraction performed well, though challenges remain in refining add-ons and usage limits extraction.

Furthermore, automating SaaS iPricing transformation has significant potential for the SaaS ecosystem. It improves consistency, reduces errors, enhances scalability, and minimizes manual effort for providers, while offering users more transparent and customizable pricing options.

Looking ahead, several lines for future research and development are proposed. Enhancements to the AI4Pricing2Yaml system could involve the mitigation of both internal and external challenges discussed in Section \ref{sec:discussion}. Additionally, exploring the use of LLM agents to interact with dynamic content could further bolster extraction capabilities. Expanding the system’s adaptability to diverse SaaS website layouts and improving its overall robustness will be crucial for its long-term success in the ever-evolving SaaS landscape.

\begin{credits}
\subsubsection*{Replicability \& Verifiability.} All artifacts from this study are available in the supplementary material \cite{LABPACK}, including a detailed results report, the Excel file and the raw outputs used in Section \ref{sec:results}, and our Information Extractor implementation.

% \subsubsection*{Acknowledgements.} This work has been partially supported by grants PID2021-126227NB-C21 and PID2021-126227NB-C22 funded by MCIN / AEI / 10.13039 / 501100011033 / FEDER, UE, and grants TED2021-131023B-C21 and TED2021-131023B-C22 funded by MCIN / AEI / 10.13039 / 501100011033 and by European Union  “NextGenerationEU”/PRTR.
\end{credits}

\bibliographystyle{splncs04}
\bibliography{references}
\end{document}